\documentclass[
  final            % use final for the camera ready runs
,letterpaper
,mathptmx]
{aipproc}

\layoutstyle{8x11single}
\usepackage{graphicx}
\usepackage{caption}
\usepackage{subcaption}

\newcommand{\rs}[1]{\textrm{\small #1}}

\begin{document}

\title{Vector meson production at the LHC}

 %\texttt{http://www.aip..org/pacs/index.html}>

\author{R.~Fiore}{
  address={ Dipartimento di Fisica, Universit\`a  della Calabria and
Istituto Nazionale di Fisica Nucleare, Gruppo collegato di Cosenza,
I-87036 Arcavacata di Rende, Cosenza, Italy}
%  email={fiore@fis.unical.it}
}

\author{L.~Jenkovszky}{
  address={Bogolyubov Institute for Theoretical Physics, %ґґ
  National Academy of Sciences of Ukraine, Kiev, 03680 Ukraine}
%  email={jenk@bitp.kiev.ua}
%  thanks={..Viiii.}  
}
\author{V.~Libov}{
  address={Deutsches Elektronen-Synchrotron, Hamburg, Germany}
%  email={vladyslav.libov@desy.de}
  %  ,altaddress={<author1 address>} % additional visiting address
}

\author{M.~V.~T.~Machado}{
  address={HEP Phenomenology Group, CEP 91501-970, Porto Alegre, RS, Brazil}
%  email={magnus@if.ufrgs.br}
}

\author{A.~Salii}{
  address={Bogolyubov Institute for Theoretical Physics, %ґґ
  National Academy of Sciences of Ukraine, Kiev, 03680 Ukraine},
%  email={saliy.andriy@gmail.com}
}

\begin{abstract}
By using a Regge-pole model for vector meson production (VMP), that successfully describes the HERA data, we analyse the connection of VMP cross sections in photon-induced reactions at HERA with those in ultra-peripheral collisions at the Large Hadron Collider (LHC).
The role of the low-energy behaviour of VMP cross sections in $\gamma p$ collisions is scrutinized.
\end{abstract}
\maketitle

%%%%%%%%%%%%%%%%%%%%%%%%%%%%%%%%%%%%%%%%%%%%
%% MAINMATTER
%%%%%%%%%%%%%%%%%%%%%%%%%%%%%%%%%%%%%%%%%%%%

\section{Ultra-peripheral collisions}\label{Int}
Following the shut-down of HERA, interest in exclusive diffractive vector meson production (VMP) has shifted to the LHC. In ultra-peripheral reactions $h_1h_2\rightarrow h_1Vh_2,$  (where $h_i$ stands for hadrons or nuclei, e.g. Pb, Au, etc.) one of the hadrons (or nuclei) emits quasi-real photons that interact with the other proton/nucleus in a similar way as in $ep$ collisions at HERA. Hence the knowledge of the 
$\gamma p\rightarrow Vp$ cross section accumulated at HERA is useful at the  LHC. The second ingredient is the photon flux emitted by the  proton (or nucleon). The importance of this class of reactions was recognized in early $70$-ies (two-photon reactions in those times) \cite{Budnev,Terazawa}. 
%In those papers, in particular the photon flux was calculated. For contemporary reviews see, e.g. Ref.~\citenum{Review}.   
In particular, in those papers the photon flux was calculated. For a contemporary review on these calculations, see for instance Ref.~\cite{Review}.

In the present paper we continue studies of VPM in ultra-peripheral collisions at the LHC
started in Ref.~\cite{TMF}, where references to previous papers can be also found. 
In particular, we make predictions for $J/\psi$ and $\psi(2S)$ productions in $pp$ scattering. 
We also extend the analysis to the lower energies for the photon-proton cross section, in order to scrutinize the $\frac{d\sigma^{pp}}{dy}$ cross section behavior at that kinematic regime.
% Also we use extended to lower energies $\gamma p$ cross sections, to scrutinize differential rapidity $\frac{d\sigma^{pp}}{dy}$ cross sections.   

The rapidity distribution of the cross section of vector meson production (VMP) in the reaction $h_1h_2\rightarrow h_1Vh_2,$ as shown in Fig.~\ref{fig:vmp_feynman}, can be written in a factorized form, i.e. it can be presented as a product of the photon flux and photon-proton cross section {\cite{Review,TMF}.% (see Eqs.~(1), (9) of \rcite{Brazil}(a))}. 

\begin{figure}[!h]
   \includegraphics[width=1.8in]{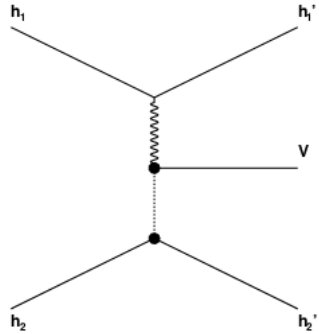}
   \caption{Feynman diagram of vector meson production in hadronic collision.}
  \label{fig:vmp_feynman}
\end{figure}

The $\gamma p \rightarrow Vp$ cross section ($V$ stands for a vector meson) depends on three variables: the total energy $W$ of the $\gamma p$ system, the squared momentum transfer $t$ and $\widetilde Q^2=Q^2+M_V^2$, where $Q^2=-q^2$ is the photon virtuality. Since, in ultraperipheral\footnote {In ultraperipheral collisions the impact parameter $b\gg R_1+R_2,$ i.e. the closest distance between the centers of the colliding particles/nuclei, $R_{1,2}$ being their radii.} collisions  photons are nearly real ($Q^2\approx 0$), the vector meson mass $M_V^2$ remains the only measure of ``hardness''. %\footnote{This might not be true for (peripheral) collisions in which the Pomeron or an $f$ reggeon is exchanged instead of the photon.}.
The $t$-dependence (the shape of the diffraction cone) is known to be nearly exponential.
It can be either integrated, or kept explicit. 
The integrated $\sigma_{\gamma p\rightarrow Vp}(\tilde Q^2, W)$ and differential    $\frac{d\sigma(t)}{dt}$ cross sections are well known from HERA measurements.

As mentioned, the differential cross section as function of rapidity can be factorized:\footnote{More precisely, the cross section can be presented as the sum of two factorized terms, depending on the photon or Pomeron emitted by the relevant proton.}
 \begin{equation}\label{eq:1}
\frac{d\sigma}{dy}^{h_1h_2\rightarrow h_1Vh_2}
=r(y)E_{\gamma_+}\frac{dN_{\gamma_+h_1}}{dE_{\gamma_+}}\sigma^{\gamma h_2\rightarrow Vh_2}(E_{\gamma_+})+
r(y)E_{\gamma_-}\frac{dN_{\gamma_-h_2}}{dE_{\gamma_-}}\sigma^{\gamma h_1\rightarrow Vh_1}(E_{\gamma_-}).
\end{equation}
Here $\frac{dN_{\gamma h}}{dE_{\gamma}}
=\frac{\alpha_{em}}{2\pi E_{\gamma}}\left[1+(1-\frac{2E_{\gamma}}{W_{pp}})^2\right]
\left(\ln\Omega-\frac{11}{6}+\frac{3}{\Omega}-\frac{3}{2\Omega^2}+\frac{1}{3\Omega^3}\right)$
is the {``equivalent''} photon flux~\cite{Review}, 
$\sigma^{\gamma h_i\rightarrow Vh_i}(E_{\gamma})$ is the total (i.e. integrated over $t$) exclusive VMP cross section (the same as at HERA \cite{ActaPol,FFJS}),
{$r(y)$ is the rapidity gap survival correction,}
and $E_{\gamma}=W^2_{\gamma p}/(2W_{pp})$ is the photon energy, with
$E_{\gamma\,\mathrm{min}}=M_V^2/(4\gamma_Lm_p),$ where $\gamma_L=W_{pp}/(2m_p)$
is the Lorentz factor (Lorentz boost of a single beam).
Furthermore,
$\Omega=1+Q_0^2/Q_\mathrm{min}^2,$ $Q_\mathrm{min}^2=\left(E_{\gamma}/\gamma_L\right)^2,$ $Q_0^2=0.71$GeV$^2,$ $x=M_Ve^{-y}/W_{pp},$ and $y=\ln(2E_{\gamma}/m_V)$. The signs $+$ or $-$ near $E_\gamma$ and $N_\gamma$ in Eq.~(\ref{eq:1}) correspond to the particular proton, to which the photon flux is attached.

For definiteness we assume that: a) the colliding particles are protons;
b) the produced vector meson $V$ is $J/\psi$ (or $\psi(2S)$), and c) the collision energy $W_{pp}=7\,$TeV.

% 
% In contrary to the previous publication \cite{TMF} here we include also a theoretical predictions for $\psi(2S)$ meson production in ultraperipheral $pp$ scattering. 

\subsection{Corrections for rapidity gap survival probabilities}\label{corrections}
The predictions may be modified by corrections due to initial and final state interactions,
alternatively called rescattering corrections.
%Calculation of these corrections is by far not unambiguous, the result, depending both on the input and on the unitarization procedure, {chosen}.
%The calculation of these corrections is by far not unambiguous and the result depends on both the physical model input and on the unitarization. %considered.
%{The better (more realistic) the input, the smaller the unitarity (rapidity gap survival probability) corrections.}
Since this is a complicated and controversial issue {\it per se}, deserving special studies beyond the scope of the present paper, here we use only familiar results from the literature:
the standard prescription is to multiply the scattering amplitude
% (cross section) 
by a factor (smaller than one), depending on energy and eventually other kinematic variables~\cite{Ryskin}.
In this work we use a constant correction coefficient $r=0.8$ (a variable one, $r(y)=0.85-0.1|y|/3$ was used in Ref.~\cite{LHCb2})}.

\section{The $\gamma p \to Vp$ cross section}
In this Section we present theoretical predictions for $J/\psi$ and $\psi(2S)$ production in $\gamma p$ scattering.
In doing so, we use the so-called Reggeometric model \cite{ActaPol}, the two-component (``soft'' and ``hard'') Pomeron model \cite{FFJS} and a model \cite{Martynov} including the low-energy region.
In the Reggeometric model we use   
\begin{equation}
\sigma_{\gamma p \to J/\psi}=A_0^2\,\frac{(W_{\gamma p}/W_0)^{4(\alpha_0-1)}}{(1+\widetilde Q^2/Q^2_0)^{2n}\left[4\alpha'\ln(W_{\gamma p}/W_0)+4\left(\frac{a}{\widetilde Q^2}+\frac{b}{2m_\mathrm{p}^2}\right)\right]}\ ,
\end{equation}
where $\widetilde Q^2=Q^2+m_V^2,$ with the parameters  
$A_0=29.8\,{{\sqrt{\rs{nb}}}/{\rs{GeV}}},\ Q_0^2=2.1\,{\rs{GeV}^2},\ n=1.37,\ \alpha_0 =1.20,\ \alpha'=0.17\,\rs{GeV}^{-2},$
$a=1.01\,{\rs{GeV}^{2}},\ b=0.44\,{\rs{GeV}^{2}},\ W_0=1\,{\rs{GeV}^{2}}$ (Ref.~\cite{FFJS}, Table II, $J/\psi$ production).

The models above, apart from $W$ and $t$, contain also dependence on the virtuality $Q^2$ and the mass of the vector meson $M_V$, relevant in extensions to the $\psi(2S)$ production cross section. %Since the masses of $J/\psi$ and $\psi(2S)$ mesons are close, we expect the parameters of the %cross section models for both cases to be similar. 
As shown in Ref.~\cite{FFJS}, to obtain the $\psi(2S)$ cross section one needs also an appropriate normalization factor, which is expected to be close to $f_{\psi(2S)}=\frac{m_{\psi(2S)}\Gamma(\psi(2S)\to e^+e^-)}{m_{J/\psi}\Gamma(J/\psi\to e^+e^-)}=0.5$\,. According to a fit of $\frac{\sigma^{\gamma p\to p+\psi(2S)}(W)}{\sigma^{\gamma p\to p+J/\psi}(W)}$ to the data \cite{psiHERAData} with a two-component Pomeron model, the value  $f_{\psi(2S)}=0.4$ is reasonable.
Thus, if the formula for the cross section $\sigma(W,Q^2,\,m_{J/\psi})$ describes $\gamma p\to J/\psi+p$ production, then $f_{\psi(2S)}\sigma(W,Q^2,\,m_{\psi(2S)})$ should describe $\gamma p\to \psi(2S)+p$ production as well.

%#### Fig2 ###############
\begin{figure}[!ht]
\includegraphics[trim = 0mm 0mm 0mm 0mm,clip,width=3.5in]{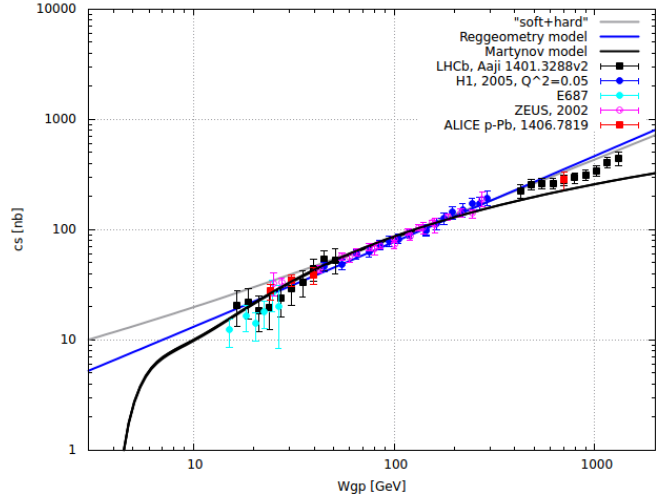}%\put(-130,100){a}%
\caption{\label{fig:csW_Jpsi} Comparison of LHCb \cite{LHCb2}, ALICE~\cite{TheALICE_2014} and HERA \cite{HERA} data on $J/\psi$ photoproduction cross section with the two component Pomeron model \cite{FFJS}, the Reggeometry model \cite{ActaPol} and the Martynov model \cite{Martynov}.}
\end{figure}
%#########################
The above mentioned models fitted to the HERA electron-proton VMP data can be applied also to the VMP in hadron-hadron scattering.
The LHCb Collaboration has recently measured ultraperipheral $J/\psi$ and $\psi(2S)$ photoproduction cross sections in $pp$-scattering (at $7\,$TeV) \cite{LHCb2}.  
From these data the $\gamma p$ cross section can be extracted. In Fig.~\ref{fig:csW_Jpsi} we compare the LHCb \cite{LHCb2}, ALICE~\cite{TheALICE_2014} and HERA \cite{HERA} data on $J/\psi$ photoproduction to the theoretical predictions. 

% As can be seen on Figs.~\ref{fig:comparison}(c)-(d), data goes steeper than our the power-like, and the Reggeometric prediction, in the range $y\in[2,\,4.5]$.
% LHCb also extracted the basic photon-proton photoproduction cross section as a function of $W_{\gamma p}$ from their rapidity data.
% The result is compared to our predictions, where the ZEUS and H1 data are also shown, Fig.~\ref{fig:comparison}(a).

%%###############################################################
\section{Rapidity distributions}\label{sec:Martynov}
\begin{figure}[!hb]
\includegraphics[trim = 0mm 1mm 0mm 0mm,clip,width=3.1in]{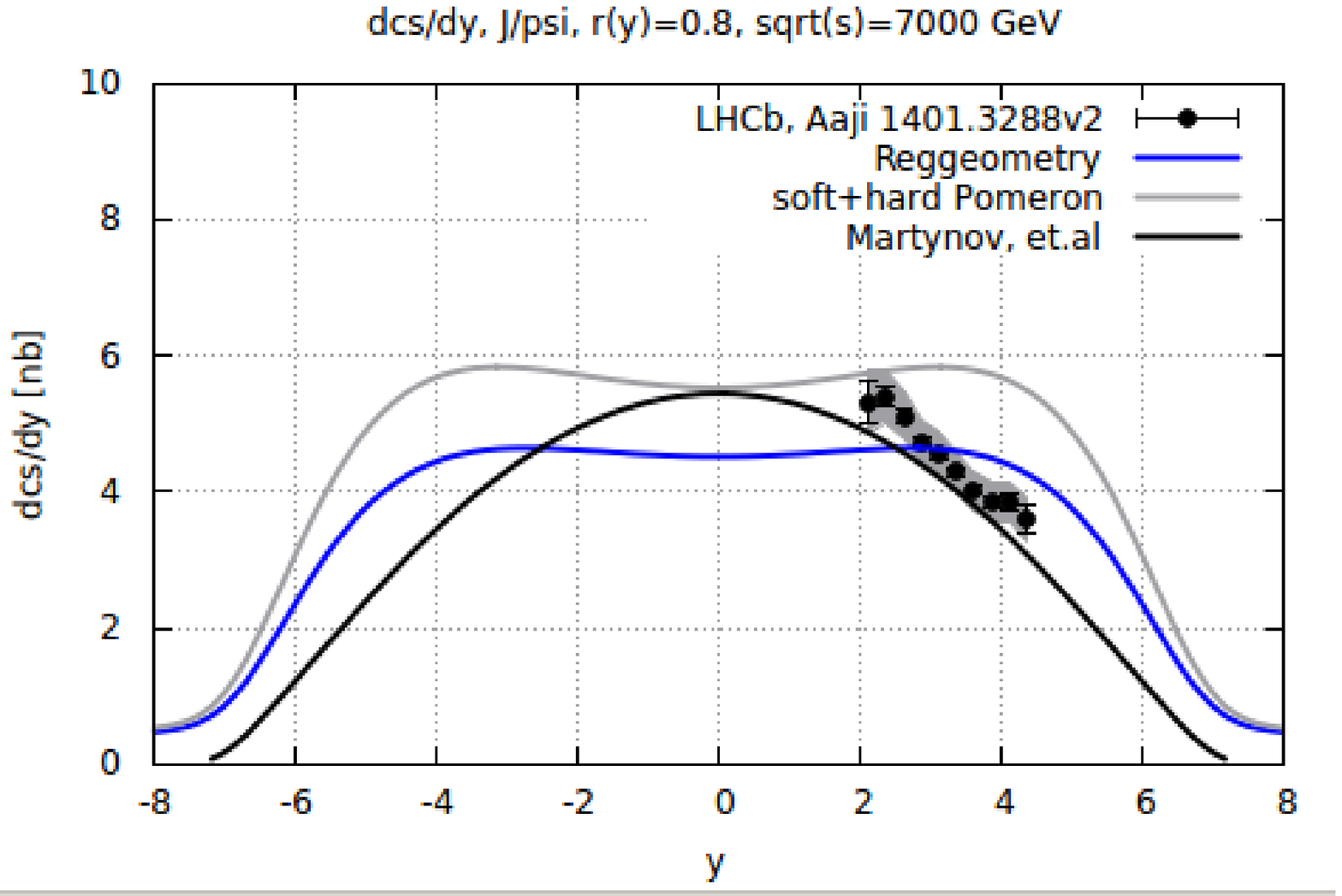}\put(-175,100){a}~
\includegraphics[trim = 1mm 1mm 0mm 0mm,clip,width=3.1in]{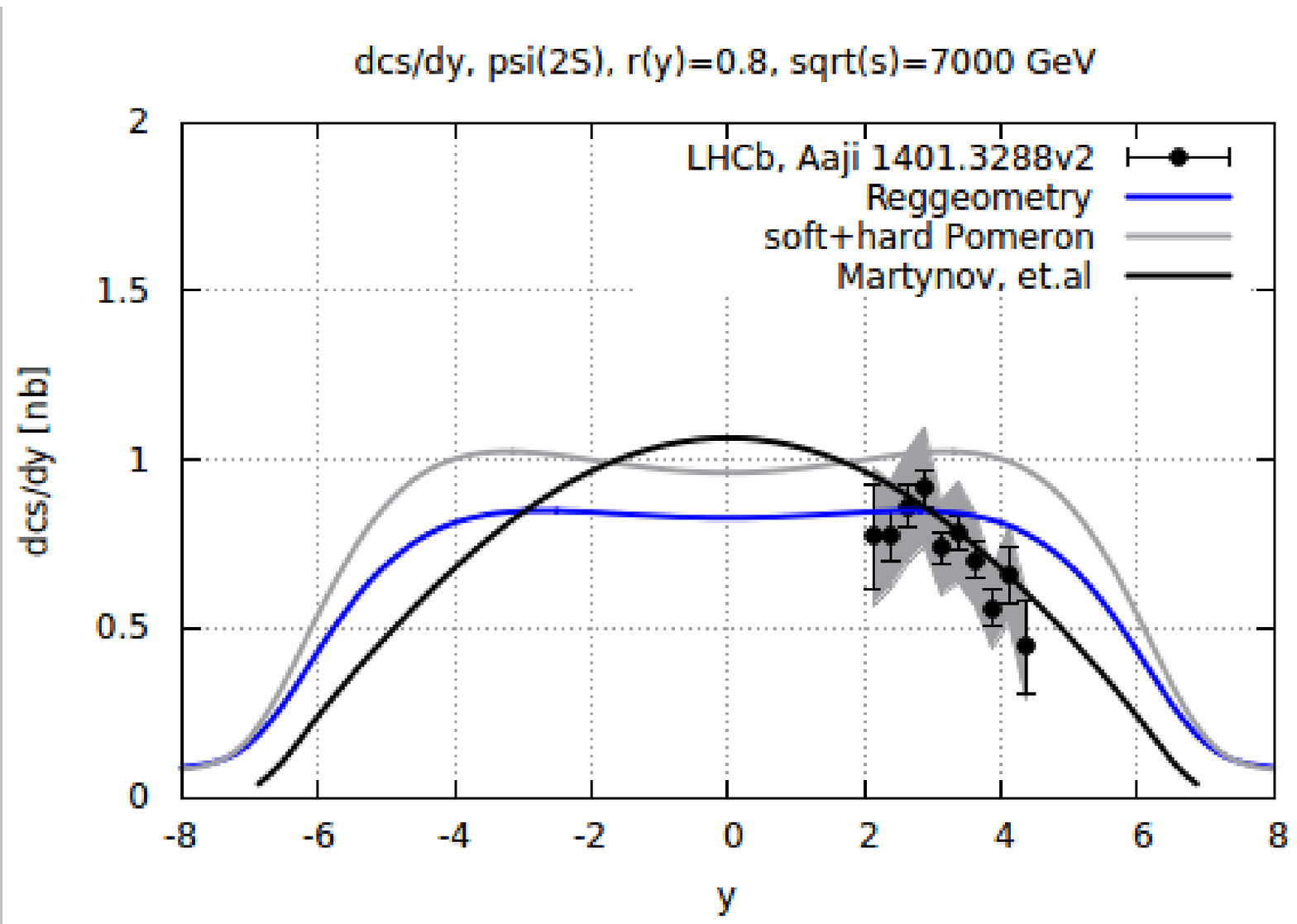}\put(-175,100){b}
\caption{Comparison of the LHCb \cite{LHCb2} data on (a)\,$J/\psi$ and (b)\,$\psi(2S)$ photoproduction cross section as a function of rapidity, $y$, with the two-component Pomeron model \cite{FFJS}, the Reggeometric model \cite{ActaPol} and that of Ref.~\cite{Martynov}.
An absorption correction $r(y)=0.8$ was applied to theoretical predictions.
%The Martynov model gives the best description of the data at low energies, see also Fig.~\ref{fig:mesons} .
}\label{fig:dcsdy}
\end{figure}

To calculate the rapidity distribution $\frac{d\sigma}{dy}^{pp\to pVp}(y)$ we use Eq.~(\ref{eq:1}), with an appropriate $\gamma p$ cross section $\sigma^{\gamma p\rightarrow Vp}(W_{\gamma p})$. 
In~Fig.~\ref{fig:dcsdy} we show the LHCb \cite{LHCb2} data together with the predictions for the $J/\psi$ and the $\psi(2S)$ differential rapidity cross sections obtained from the Regge model \cite{ActaPol}, the two-component Pomeron model \cite{FFJS} and that of Ref.~\cite{Martynov}. The rapidity gap survival factor $r(y)=0.8$\, was used.
%#### Fig3 ###############
%######################### 

%, to transform the integrated photoproduction cross section $\sigma^{\gamma p\rightarrow J/\psi+p}(W_{\gamma p})$ into differential $\frac{d\sigma}{dy}^{pp=pp+J/\psi}(y)$.
The energy of the $\gamma p$ system $W_{\gamma p}$ is related to rapidity $y$ via $W^{\pm}_{\gamma p}=\sqrt{M_{J/\psi} W_{pp}\, e^{\pm y}}$ (the choice of the sign depends on the  propagation direction of $\gamma$). 
Hence, the differential rapidity cross section $\frac{d\sigma}{dy}$,  in the range $y\in[2,\, 4.5]$  at energy $W_{pp}=7\,$TeV, needs the knowledge of the integrated cross section  $\sigma(W_{\gamma p})$ in the range $W_{\gamma p}\in [15.5,\;54]\cup[400,1397]\,$GeV.
Fig.~\ref{fig:dcsdy}(a) and \ref{fig:dcsdy}(b) show how sensitive the differential $\frac{d\sigma}{dy}$ cross section predictions are to the choice of $\sigma(W_{\gamma p})$ cross sections (see  also Fig.~\ref{fig:csW_Jpsi}).

% The saturation model of Gay Ducati, Griep \& Machado \rcite{Brazil}(f) and the saturation model of Motyka \& Watt, \rcite{MotykaWatt}, describe the LHCb data pretty well (see \cite{LHCb2}, Fig.~7).
% Figures~\ref{fig:comparison}\,(b), (c), and (d) show that the Reggeometry model and the power-law models do not fit the LHCb data well  (in the region of $y\in[2,\,4.5]$), however the  model \cite{Martynov} fits the data better(as well as that of Gay Ducati, Griep \& Machado and the Motyka \& Watt models). One distinction between the Reggeometric model and power-like models on one hand, and the model \cite{Martynov} on the other hand is that the latter properly describe the low-energy (below $30\,$GeV) cross sections.
% Similarly the Motyka \& Watt model also better fits the low-energy data than power-law models, see Fig.~4 of \rcite{MotykaWatt} and Fig.~2 of \rcite{Martynov}.
% 
% Another interesting point to remark is that the Motyka \& Watt model and the model \cite{Martynov} goes below the power-like models cross sections not only at low energies (below $30\r{\,GeV}$) but also at high energies (above $300\r{\,GeV}$). 

% However, with the current data set we cannot say 
% whether the high-energy data should obey power-like behaviour, or they rise like a logarithm with increasing energy. 

The curve of the model \cite{Martynov} for $J/\psi$ (and its extension for $\psi(2S)$) gives better description of the differential rapidity cross section than the Reggeometric and the 
two-component Pomeron models, but it seems to underestimate the $J/\psi$ data (see Fig.~\ref{fig:dcsdy}(a)). This may result  from underestimation of the $\gamma p$ cross sections by the model \cite{Martynov} at higher energies (see Fig.~\ref{fig:csW_Jpsi}). To properly describe the rapidity distribution of VMP cross section in $pp$ scattering, we need to correctly describe the $\gamma p$ VMP cross section in the whole energy region. Each $\gamma p$ energy range corresponds to its particular rapidity range.
%GayGriepMachado MotykaWatt
%...
\vspace{-2mm}
%%%#############################################################################
%\section{Open problems, prospects}
\section{Conclusions and prospects}
\paragraph{Summary of the reported results}
We have compared several theoretical models: the Reggeometric, the power-like, and the Regge-pole model of Ref.~\cite{Martynov} with the recent LHCb data on $J/\psi$ and $\psi(2S)$ photoproduction in ultraperipheral $pp$ collisions at LHCb \cite{LHCb2}. From the integrated $\gamma p$ cross sections {\small $\sigma(W_{\gamma p})$}, obtained at HERA, we have calculated the rapidity distribution of differential $\frac{d\sigma}{dy}^{pp\to pVp}(y)$ cross section. %The model \cite{Martynov} gives the best description of the data, although after corrections %it seems to slightly underestimate the data. 
\paragraph{Prospects/problems}
In the near future the following items are on the agenda: 
\begin{enumerate}
\item
%large amount of data on the production of various (heavy and light) vector meson
%is being collected by the experimentalist at the LHC in nucleon and nuclear collisions.
%We remind that the HERA provides data only on proton (not nuclear) target. 
A feasible formalism relating nucleon and nuclear reactions should be elaborated within the Glauber theory of multiple scattering. Some work in this direction has already been done in Refs.~\cite{Review}\,a) and  \cite{Kopel}. 
\item HERA had provided rich information on the $Q^2$ and $t$ dependence of VMP production, and sofisticated models exist (see, for example, Ref.~\cite{FFJS} and references therein) reproducing this rich and non-trivial dependence. This is not known yet at the LHC: in nuclear collisions the $Q^2$  (by kinematics, $Q^2$ in nuclear collisions is limited to less than $0.5\,$GeV$^2$) and the $t$ dependence are practically unknown. We hope that the experimental situation will improve. 

 \item The $f$ trajectory may append/replace the Pomeron exchange, and similarly the photon flux may be appended by the flux of $\omega$'s and of the Odderon trajectories, opening new channels and thus making the picture more complicated, but, at the same time, more interesting.    
 \end{enumerate}
\vspace{-2mm}
 
%%%%%%%%%%%%%%%%%%%%%%%%%%%%%%%%%%%%%%%%%%%%%%%%
%% BACKMATTER
%%%%%%%%%%%%%%%%%%%%%%%%%%%%%%%%%%%%%%%%%%%%%%%%
\begin{theacknowledgments}
L.L. J. thanks the Organizers of this Conference for hospitality and financial support.
\end{theacknowledgments}
\vspace{-2mm}



\bibliographystyle{aipproc}   % if natbib is available

%\bibliography{sample}

\begin{thebibliography}{9}
%## 1 ###
\bibitem{Budnev} %\alt{V.E. Balakin, V.M. Budnev, I.F. Ginzburg, Pis'ma v Zhetf, {\bf 25}, 559 (1970) (JETP Lett. {\bf 11}, 338 (1970).}
  V.~E.~Balakin, V.~M.~Budnev and I.~F.~Ginzburg,
  %``Possible experiment of hadron production by two photons from threshold to extremely high energies,''
  Pisma Zh.\ Eksp.\ Teor.\ Fiz.\  {\bf 11}, 559 (1970).
  %%CITATION = ZFPRA,11,559;%%
  %16 citations counted in INSPIRE as of 12 Oct 2014

%## 2 ###
\bibitem{Terazawa}% \alt{Stanley J. Brodsky, Toichiro Kinoshita and Hidezumi Terazawa, Phys. Rev. Letters, {\bf 25} 972 (1970).} 
  S.~J.~Brodsky, T.~Kinoshita and H.~Terazawa,
  %``Dominant colliding beam cross-sections at high-energies,''
  Phys.\ Rev.\ Lett.\  {\bf 25}, 972 (1970).
  %%CITATION = PRLTA,25,972;%%
  %169 citations counted in INSPIRE as of 12 Oct 2014

%## 3 ###
\bibitem{Review}
 G.~Baur {\it et al.}, Rev. Mod. Phys. {\bf 50} 26; (1978);
 G.~Baur {\it et al.}, Phys. Rept. {\bf 364}, 359 (2002), hep-ph/0112211; 
 K.~Hencken {\it et al.}, Phys. Rept. {\bf 458}, 1 (2008), arXiv:0706.3356.


%## 4 ###
\bibitem{TMF} R.~Fiore {\it et al.}, {\it ``Vector meson production in ultra-peripheral collisions at the LHC''}, Theor. and Mathematical Physics, in press, arXiv:1408.0530.
% % \bibitem{Brazil} a)V.~P.~Goncalves and M.~M.~Machado, Phys. Rev. {\bf D85}, 054019 (2012), arXiv:1112.3500;\\
% % b)%{\it ibid.} 
% % V.~P.~Goncalves and M.~M.~Machado, Phys. Rev. 
% % {\bf C84}, 011902 (2011), arXiv:1106.3036;\\
% % c)%{\it ibid.} 
% % V.~P.~Goncalves and M.~M.~Machado, Phys. Rev. 
% % {\bf C80} 054901 (2009), arXiv:0907.4123;\\
% % d)V.~P.~Goncalves and M.~M.~Machado, Eur. Phys. J. {\bf C72}, 2231 (2012), arXiv:1207.5273;\\
% % e)V.~P.~Gonsalves and W.~K.~Suter, Eur. Phys. J. {\bf A47}, 117 (2011), arXiv: 1004.1952;\\
% % f)V.~P.~Goncalves, Nucl. Phys. {\bf A902}, 32 (2013), arXiv:1211.1207;\\
% % g)M.~B.~Gay Ducati {\it et al.}, %, M.~T.~Griep, and M.V.T. Machado,
% % Phys. Rev. {\bf D88}, 017504 (2013), arXiv:1305.4611;\\
% % h)G. Sampaio dos Santos, M.~V.~T.~Machado, 
% % arXiv:1407.4148;\\
% % i)$\;$% %## 8 ###
% % %\bibitem{MotykaWatt}
% % L.~Motyka and G.~Watt, %Exclusive photoproduction at the Fermilab Tevatron and CERN LHC within the dipole picture,
% % Phys. Rev. {\bf D78}, 014023 (2008), arXiv:0805.2113.

%## 5 ###
\bibitem{ActaPol}
 S.~Fazio {\it et al.},
%, R.~Fiore, A.~Lavorini, L.~Jenkovszky and A.~Salii,
  %``Reggeometry of deeply virtual Compton scattering (DVCS) and exclusive vector meson production (VMP) at HERA,''
  Acta Phys.\ Polon.\ B{\bf 44}, 1333 (2013),
  arXiv:1304.1891. % [hep-ph]].
\bibitem{FFJS}  
  S.~Fazio {\it et al.}, %  R.~Fiore, L.~Jenkovszky and A.~Salii,
  %``Unifying "soft" and "hard" diffractive exclusive vector meson production and deeply virtual Compton scattering,''
  Phys.\ Rev.\ {\bf D90}, 016007 (2014),
  arXiv:1312.5683. % [hep-ph]].

%## 6 ###
\bibitem{Ryskin}
V.~A.~Khoze {\it et al.}, % A.~D.~Martin, and M.~G.~Ryskin,
Eur. Phys. J. {\bf C24}, 459 (2002), hep-ph/0201301;
S.~P.~Jones {\it et al.}, % A.~D.~Martin, M.~G.~Ryskin, and T.~Teubner,
  %``Probes of the small $x$ gluon via exclusive $J/\psi$ and $\Upsilon$ production at HERA and the LHC,''
  JHEP {\bf 1311}, 085 (2013), arXiv:1307.7099;
S.~P.~Jones {\it et al.}, % A.~D.~Martin, M.~G.~Ryskin, and T.~Teubner,
J. Phys. {\bf G41}, 055009 (2014), arXiv:1312.6795.
  
% \bibitem{LHCb1} LHCb Collab., R. Aaij {\it et al.}, J. Phys. {\bf G40}, 045001 (2013), arXiv:1301.7084.
%## 7 ###
\bibitem{LHCb2} [LHCb Collaboration], R. Aaij {\it et al.}, J. Phys. {\bf G41}, 055002 (2014), arXiv:1401.3288;
{\it ibid.}  {\bf G40}, 045001 (2013), arXiv:1301.7084.
%## 8 ###
\bibitem{Martynov} 
  E.~Martynov, E.~Predazzi and A.~Prokudin,
  %``Photoproduction of vector mesons in the soft dipole pomeron model,''
  Phys.\ Rev.\ {\bf D67}, 074023 (2003),
%  \Journal{\PRD}{67}{074023}{2003}.
  hep-ph/0207272. 
  
 %%% Data %%%%%%%%%%%%%%%%%%%%%%%%%%%%%%%%%%%%%%%%%%%%%%%%%%%%%%%%%%%%%%%%%%%%%%
%## 9 ###
  \bibitem{psiHERAData}
%  \bibitem{ZEUSprelim}
   [ZEUS Collaboration], 
   {\it ``Measurement of the cross-section ratio $\mathit{\sigma_{\psi(2S)}/\sigma_{J/\psi(1S)}}$ in deep-inelastic exclusive $ep$ scattering at HERA''} (prelim.) (2014);
%
  %  \bibitem{H1_1998} %{Adloff:1997yv}
  [H1 Collaboration], C.~Adloff {\it et al.} 
  %``Photoproduction of psi (2S) mesons at HERA,''
  Phys.\ Lett.\ B {\bf 421} (1998) 385
  [hep-ex/9711012];
%  
% %\bibitem{H1_2002} %{Adloff:2002re}
%   C.~Adloff {\it et al.}  [H1 Collaboration],
%   %``Diffractive photoproduction of psi(2S) mesons at HERA,''
%   Phys.\ Lett.\ B 
  {\it ibid.}
  {\bf 541} (2002) 251
  [hep-ex/0205107];
%
 %\bibitem{H1_1999} %{Adloff:1999zs}
  [H1 Collaboration], C.~Adloff {\it et al.}, 
  %``Charmonium production in deep inelastic scattering at HERA,''
  Eur.\ Phys.\ J.\ C {\bf 10} (1999) 373
  [hep-ex/9903008].

  \bibitem{TheALICE_2014}
   [ALICE Collaboration], B.~Abelev {\it et al.}, 
  %``Exclusive $J/\psi$ photoproduction off protons in ultra-peripheral p-Pb collisions at $\sqrt{s_{NN}}$=5.02 TeV,''
  arXiv:1406.7819 [nucl-ex].
  %%CITATION = ARXIV:1406.7819;%%
  
%## 10 ###
\bibitem{HERA}%
[ZEUS Collaboration], S.~Chekanov {\it et al.},
  %``Exclusive photoproduction of J / psi mesons at HERA,''
  Eur.\ Phys.\ J.\ {\bf C24}, 345 (2002),
  hep-ex/0201043;
[H1 Collaboration],  A.~Aktas {\it et al.},
  %``Elastic J/psi production at HERA,''
  Eur.\ Phys.\ J.\ {\bf C46}, 585 (2006),
  hep-ex/0510016;
[E687 Collaboration], P.~Frabetti {\it et al.}, Phys. Lett. B {\bf316}, 197 (1993).

%## 11 ###
\bibitem{Kopel} J.~H\"ufner, B.~Kopeliovich and J.~Nemchik, Phys. Lett. B {\bf 383}, 362 (1996).

%### Other not included ##########################################


 
\end{thebibliography}

% \IfFileExists{\jobname.bbl}{}
%  {\typeout{}
%   \typeout{******************************************}
%   \typeout{** Please run "bibtex \jobname" to optain}
%   \typeout{** the bibliography and then re-run LaTeX}
%   \typeout{** twice to fix the references!}
%   \typeout{******************************************}
%   \typeout{}
%  }

%%%%%%%%%%%%%%%%%%%%%%%%%%%%%%%%%%%%%%%%%%%
%% The following lines show an example how to produce a bibliography
%% without the help of the BibTeX program. This could be used instead
%% of the above.
%%%%%%%%%%%%%%%%%%%%%%%%%%%%%%%%%%%%%%%%%%%

\end{document}